\documentclass[aps,physrev,superscriptaddress,showkeys,floatfix,twocolumn]{revtex4-2}
\usepackage[colorlinks=true, allcolors=blue]{hyperref}
\usepackage{siunitx}	% for SI units
\usepackage{braket}
\usepackage{amsmath}
\usepackage{graphicx}
\usepackage{subcaption}
\usepackage{makecell}
\graphicspath{{images/}}

\usepackage[commandnameprefix=ifneeded]{changes}
\definechangesauthor[name={Shikha}, color=blue]{SR}

\begin{document}

% Use the \preprint command to place your local institutional report
% number in the upper righthand corner of the title page in preprint mode.
% Multiple \preprint commands are allowed.
% Use the 'preprintnumbers' class option to override journal defaults
% to display numbers if necessary
%\preprint{}

%Title of paper
% I hate this paper
\title{Charge radii of Cl isotopes from x-ray spectroscopy of muonic atoms}

% repeat the \author .. \affiliation  etc. as needed
% \email, \thanks, \homepage, \altaffiliation all apply to the current
% author. Explanatory text should go in the []'s, actual e-mail
% address or url should go in the {}'s for \email and \homepage.
% Please use the appropriate macro foreach each type of information

% \affiliation command applies to all authors since the last
% \affiliation command. The \affiliation command should follow the
% other information
% \affiliation can be followed by \email, \homepage, \thanks as well.

\author{K.A.~Beyer}
\affiliation{Max-Planck-Institut für Kernphysik, Heidelberg, Germany}

\author{T.E.~Cocolios}
\affiliation{KU Leuven, Instituut voor Kern- en Stralingsfysica, Leuven, Belgium}

\author{C.~Costache}
\affiliation{Horia Hulubei National Institute for R\&D in Physics and Nuclear Engineering, Bucharest, Romania}

\author{P.~Demol}
\affiliation{Université Libre de Bruxelles, Institut d’Astronomie et d’Astrophysique, Brussels, Belgium}
\affiliation{Brussels Laboratory of the Universe-BLU-ULB, Brussels, Belgium}

\author{M.~Deseyn}
\affiliation{KU Leuven, Instituut voor Kern- en Stralingsfysica, Leuven, Belgium}

\author{A.~Doinaki}
\affiliation{PSI Center for Neutron and Muon Sciences, Villigen, Switzerland}
\affiliation{ETH Zürich, Institute for Particle Physics and Astrophysics, Zürich, Switzerland}

\author{O.~Eizenberg}
\affiliation{The Helen Diller Quantum Center, Department of Physics, Technion-Israel Institute of Technology, Haifa, Israel}

\author{M.~Gorchtein}
\affiliation{Institut f\"ur Kernphysik, Johannes Gutenberg-Universit\"at Mainz, Mainz, Germany}
\affiliation{PRISMA$^{++}$ Cluster of Excellence, Johannes Gutenberg-Universit\"at Mainz, Mainz, Germany}

\author{M.~Heines}
\email[]{Corresponding author: michael.heines@kuleuven.be\\ Current affiliation: PSI center for Neutron and Muon Sciences.}
\affiliation{KU Leuven, Instituut voor Kern- en Stralingsfysica, Leuven, Belgium}

\author{A.~Herzáň}
\affiliation{Institute of Physics, Slovak Academy of Sciences, Bratislava, Slovakia}

\author{P.~Indelicato}
\affiliation{Laboratoire Kastler Brossel, Sorbonne Université, CNRS, ENS-PSL Research University, Collège de France, Paris, France}

\author{K.~Kirch}
\affiliation{PSI Center for Neutron and Muon Sciences, Villigen, Switzerland}
\affiliation{ETH Zürich, Institute for Particle Physics and Astrophysics, Zürich, Switzerland}

\author{A.~Knecht}
\affiliation{PSI Center for Neutron and Muon Sciences, Villigen, Switzerland}

\author{R.~Lic{\u a}}
\affiliation{Horia Hulubei National Institute for R\&D in Physics and Nuclear Engineering, Bucharest, Romania}

\author{V.~Matousek}
\affiliation{Institute of Physics, Slovak Academy of Sciences, Bratislava, Slovakia}

\author{E.A.~Maugeri}
\affiliation{PSI Center for Nuclear Engineering and Sciences, Villigen, Switzerland}

\author{B.~Ohayon}
\affiliation{The Helen Diller Quantum Center, Department of Physics, Technion-Israel Institute of Technology, Haifa, Israel}

\author{N.S.~Oreshkina}
\affiliation{Max-Planck-Institut für Kernphysik, Heidelberg, Germany}

\author{W.W.M.M.~Phyo}
\affiliation{KU Leuven, Instituut voor Kern- en Stralingsfysica, Leuven, Belgium}

\author{R.~Pohl}
\affiliation{Institut f\"ur Physik, QUANTUM, Johannes Gutenberg-Universit\"at Mainz, Mainz, Germany}
\affiliation{PRISMA$^{++}$ Cluster of Excellence, Johannes Gutenberg-Universit\"at Mainz, Mainz, Germany}
\affiliation{SFB1660, Institut f\"ur Kernphysik and Institut f\"ur Physik, Johannes Gutenberg-Universit\"at Mainz, Mainz, Germany}

\author{S.~Rathi}
\affiliation{The Helen Diller Quantum Center, Department of Physics, Technion-Israel Institute of Technology, Haifa, Israel}

\author{W.~Ryssens}
\affiliation{Université Libre de Bruxelles, Institut d’Astronomie et d’Astrophysique, Brussels, Belgium}
\affiliation{Brussels Laboratory of the Universe-BLU-ULB, Brussels, Belgium}

\author{K.~von Schoeler}
\affiliation{ETH Zürich, Institute for Particle Physics and Astrophysics, Zürich, Switzerland}

\author{A.~Turturica}
\affiliation{Horia Hulubei National Institute for R\&D in Physics and Nuclear Engineering, Bucharest, Romania}

\author{I.A.~Valuev}
\affiliation{Max-Planck-Institut für Kernphysik, Heidelberg, Germany}

\author{S.M.~Vogiatzi}
\affiliation{KU Leuven, Instituut voor Kern- en Stralingsfysica, Leuven, Belgium}
%\affiliation{Institut f\"ur Physik, QUANTUM, Johannes Gutenberg-Universit\"at Mainz, Mainz, Germany}
\affiliation{SFB1660, Institut f\"ur Kernphysik and Institut f\"ur Physik, Johannes Gutenberg-Universit\"at Mainz, Mainz, Germany}

\author{F.~Wauters}
\affiliation{Institut f\"ur Kernphysik, Johannes Gutenberg-Universit\"at Mainz, Mainz, Germany}
\affiliation{PRISMA$^{++}$ Cluster of Excellence, Johannes Gutenberg-Universit\"at Mainz, Mainz, Germany}

\author{A.~Zendour}
\affiliation{PSI Center for Neutron and Muon Sciences, Villigen, Switzerland}
\affiliation{ETH Zürich, Institute for Particle Physics and Astrophysics, Zürich, Switzerland}

%\email[]{Your e-mail address}
%\homepage[]{Your web page}
%\thanks{}
%\altaffiliation{}

%Collaboration name if desired (requires use of superscriptaddress
%option in \documentclass). \noaffiliation is required (may also be
%used with the \author command).
%\collaboration can be followed by \email, \homepage, \thanks as well.
%\collaboration{}
%\noaffiliation

\date{\today}

\begin{abstract}
    Nuclear charge radii are vital for nuclear and atomic physics, the determination of fundamental constants, and searches for new physics. Muonic atoms, where a single negative muon orbits a nucleus, are sensitive tools for determining nuclear radii due to the large wavefunction overlap of the muon and nucleus. Here we report on a new measurement of the $2, 3, 4p\to1s$ x-ray energies in muonic $^{35,37}$Cl with uncertainties reaching 18 ppm. By employing a large-scale germanium detector array, it was possible to extract these energies from a high statistics dataset using highly enriched samples of only a few tens of milligrams. Combining these results with state-of-the-art atomic and nuclear theory input, the charge radii of the stable chlorine isotopes were determined to be $R(^{35}\text{Cl}) = 3.3333(23)~\si{\femto\meter}$ and $R(^{37}\text{Cl}) = 3.3444(23)~\si{\femto\meter}$. This is an order of magnitude more precise and significantly different from previously tabulated values. Our new values solve a discrepancy observed for the charge radius difference in mirror nuclei, agreeing with the overall global trend. The charge radius difference $\delta \langle r^2 \rangle (^{37}\text{Cl} - {^{35}\text{Cl}}) = -0.0776(64)~\si{\femto\meter\squared}$ we extract is 25 times more precise than the previous values. This precision is crucial for establishing reference values for future laser spectroscopy measurements of radioactive isotopes.

    %High-precision absolute nuclear charge radii are fundamental observables and provide essential inputs for a broad range of studies in nuclear, atomic, and fundamental physics. Muonic atoms offer one of the most sensitive tools for determining such radii, due to the large wavefunction overlap of the muon and nucleus. This work presents a fully modern pipeline for extracting charge radii from muonic x rays in the medium-mass region, demonstrated using $^{35, 37}$Cl. This approach included a state-of-the-art experiment as well as a rigorous and internally consistent theoretical framework tailored to this purpose. We report on the first extraction of absolute charge radii from muonic x rays with below per-mille precision in 30 years, showing control over systematic uncertainties and improved precision on the QED calculations in medium-mass nuclei. The extracted radii differ significantly from literature values from electron scattering, while achieving substantially higher precision. Their improved agreement with a global fit of mirror nuclei across the nuclear chart instills confidence in the new values. Additionally, the correlation among different observables between the two isotopes was exploited to achieve even more precise differences in mean-square charge radius. This precision is crucial for establishing benchmark values for laser spectroscopy of radioactive isotopes.
\end{abstract}

% insert suggested keywords - APS authors don't need to do this
%\maketitle must follow title, authors, abstract, and keywords
\maketitle

% body of paper here - Use proper section commands
\textit{Introduction -}
    The size of the atomic nucleus is a fundamental property, most often expressed as the root-mean-square (RMS) nuclear charge radius. It is a sensitive probe of nuclear structure, revealing effects such as shell closures~\cite{koszorus2021charge, gustafsson2025Snlaser}, nuclear pairing effects~\cite{degroote2020measurement}, and nuclear deformations~\cite{marsh2018Hglaser}. In this context, nuclear charge radii have been broadly studied throughout the nuclear landscape~\cite{angeli2013table}. While many studies probe changes in radii through isotope shifts using laser spectroscopy~\cite{yang2023laser}, they require anchoring to reference nuclear charge radii. Traditionally, such references were obtained for stable isotopes using muonic x-ray spectroscopy~\cite{fricke2004nuclear} and elastic electron scattering~\cite{de1987nuclear}.

    Recently, several physics cases have emerged in which the absolute charge radius constitutes a leading source of systematic error. The charge radius is, for example, critical for the determination of the V\textsubscript{ud} element of the Cabibbo-Kobayashi-Maskawa (CKM) quark mixing matrix~\cite{plattner2023nuclear, seng2024data, 2025-CKM, 2025-IsospinCKM}. It is also used as input in studies on the equation of state for nuclear matter to describe the properties of neutron stars~\cite{ding2023investigation, pineda2021charge}, the interpretation of atomic parity violation experiments~\cite{wansbeek2012charge}, and high-precision measurements in electronic atoms~\cite{71d6-w383}. Moreover, absolute charge radii are crucial ingredients in the parameter adjustment of several nuclear structure models: \textit{ab initio}~\cite{hu2022ab, ekstrom2015accurate, arthuis2024neutron}, energy density functionals (EDF)~\cite{chabanat1998skyrme, grams2025skyrme, reinhard2024extended}, and various types of phenomenological models~\cite{royer2009liquid, duflo1994phenomenological}. Finally, radius difference inputs are used in the determination of mass and field shift factors through King plots~\cite{king2013isotope}, which are critical for the radius extraction in laser spectroscopy studies~\cite{sahoo2024recent}. A lack of reference radii often leads to large systematic error bands that limit the precision of extracted differential mean-square charge radii of isotopes far from stability~\cite{heylen2016changes, cheal2012laser}.

    Since the muon is 207 times heavier than the electron, its atomic orbitals lie proportionally closer to the nucleus. As a result, effects induced by the finite nuclear size are enhanced by a factor $207^3\approx10^7$ in muonic atoms with respect to their electronic counterparts. Therefore, muonic atoms provide a sensitive probe for the spatial extent of the nuclear charge distribution. The highest sensitivity is achieved for transitions involving the $1s$ orbital, which has the largest overlap with the nucleus. The energies of muonic x rays towards the $1s$ level range from a few $\si{\kilo\eV}$ in light elements up to $\sim10\,\si{\mega\eV}$ in the heaviest systems.

    The radii of the stable isotopes of chlorine are currently deduced from elastic electron scattering measurements with limited precision~\cite{briscoe1980elastic}. Moreover, they have recently been found to deviate from their expected values based on the global behavior of mirror nuclei~\cite{ohayon2024critical}. An updated high-precision measurement of the chlorine charge radii could help resolve this discrepancy or, alternatively, reveal interesting effects if confirmed. Additionally, such measurements would greatly benefit the calibration of future laser spectroscopy studies on radioactive isotopes by means of a King plot. Here, the evolution of radii across the neutron $1f_{7/2}$ orbital in the proximity of calcium is a major topic of interest~\cite{blaum2008nuclear, koszorus2021charge, garcia2016unexpectedly, bai2025charge}.

    There has been a revival of interest in muonic atoms, starting with the lightest elements H, D, and He, which could be probed by laser spectroscopy~\cite{pohl2010size, pohl2016laser, krauth2021measuring, schuhmann2025helion}. Since then, additional work in heavier systems has been initiated, primarily aimed at measuring long-lived radioactive nuclei~\cite{antognini2020measurement, vogiatzi2023studies, sun2025208, vogiatzi2023studies}. The renewed interest prompted experimental efforts to probe smaller samples with higher sensitivity~\cite{adamczak2023muonic, antwis2025comparative} and theoretical efforts to include previously omitted corrections~\cite{sun2025208, gorchtein2025hitchhiker, 2026-Hybrid}. This revitalization also prompted a reevaluation of the existing Zr, Sn, and Pb data~\cite{sun2025208, beyer2025relativistic}, and a low-statistics measurement on Pd isotopes~\cite{saito2025muonic}.

    In this letter, we report on the extraction of the absolute charge radii of $^{35}$Cl and $^{37}$Cl by means of muonic x-ray spectroscopy. We combine new experimental data taken with a state-of-the-art experimental setup and a recently developed theoretical framework tailored for medium-mass muonic atoms. Additionally, the advancements in muon beams and the use of a large germanium detector array allowed for high statistics measurements of highly isotopically pure samples with target masses of only a few tens of milligrams of the element of interest. Apart from the increased statistics, a large detector array allows for better control of detector-specific systematic uncertainties.
    
    %Since then, additional work in heavier systems has been initiated~\cite{saito2025muonic, antognini2020measurement, vogiatzi2023studies, sun2025208, vogiatzi2023studies}, although only the work on Pd has published a charge radius, and their results are limited by low statistics and calibration systematics. Additionally, new experimental and theoretical efforts have been invested to probe smaller samples with higher sensitivity~\cite{adamczak2023muonic, antwis2025comparative} and provide higher accuracy in charge radii thanks to a better treatment of systematic effects for mid-size and heavy nuclei, including previously omitted theoretical corrections~\cite{sun2025208, gorchtein2025hitchhiker}. In this letter, we report on the extraction of absolute charge radii of $^{35,37}$Cl by means of muonic x-ray spectroscopy, combining the latest experimental and theoretical developments in this field. This opens a new era of absolute charge radii determination across the full nuclear landscape with higher accuracy and sensitivity. 

\textit{Experiment - }
    The muonic x-ray measurements presented in this work were performed at the $\pi$E1 beamline of the high-intensity proton accelerator facility~\cite{grillenberger2021high} at the Paul Scherrer Institute (PSI) in Switzerland. Here a continuous muon rate on the order of a few $10~\si{\kilo\hertz}$ is available at the momenta of interest. For these measurements, enriched targets containing $^{35}$Cl (200\,$\si{\milli\gram}$ of NaCl, $99.32(5)\%$ isotopic purity) and $^{37}$Cl (70\,$\si{\milli\gram}$ of AgCl, $99.32(5)\%$ isotopic purity) were provided by the Institute Laue-Langevin (ILL) and Argonne National Laboratory (ANL), respectively. The momentum of the incident muon beam was scanned around $30\,\si{\mega\eV/c}$ to maximize the signal for each target, followed by a long continuous measurement (five hours for $^{35}$Cl and ten hours for $^{37}$Cl).

    The muonic x rays were measured with the GIANT high-purity germanium detection array~\cite{gerchow2023germanium, adamczak2023muonic}, consisting of 14 detectors, including a TIGRESS-type clover~\cite{scraggs2005tigress} and a Miniball cluster detector~\cite{warr2013miniball} (total of 19 crystals). This array reached an efficiency of $3\%$ and a full width at half maximum (FWHM) resolution of $2.6\,\si{\kilo\eV}$ at $1332\,\si{\kilo\eV}$. The setup was complemented with plastic scintillators for coincidence and veto logic, which can be used to gate on incoming muons and limit the Bremsstrahlung induced by Michel electrons (arising from the decay of muons). 
    
    The data were recorded using 14-bit SIS3316 digitizers~\cite{Sis3316} with a sampling rate of $250~\si{\mega\hertz}$ and digital signal processing through trapezoidal filters for the energy determination. In order to improve the timing resolution, raw traces with a length of $1.2~\si{\micro\second}$ (of which $0.8~\si{\micro\second}$ pre-trigger) were digitized for each germanium detector waveform for offline timing optimization. These traces were then used for extrapolated leading edge timing~\cite{skawran2021development}, leading to a time resolution (FWHM) of $\sim14~\si{\nano\second}$ at approximately $700~\si{\kilo\eV}$ for scintillator-germanium time differences. This allows for strong background suppression based on time cuts relative to incoming muons. In order to provide a continuous source of calibration for the detector array, several calibration sources ($^{60}$Co, $^{110m}$Ag, $^{133}$Ba) were placed in the vicinity of the target position.

\begin{figure}[hb]
        \centering
        \includegraphics[width=1.0\linewidth]{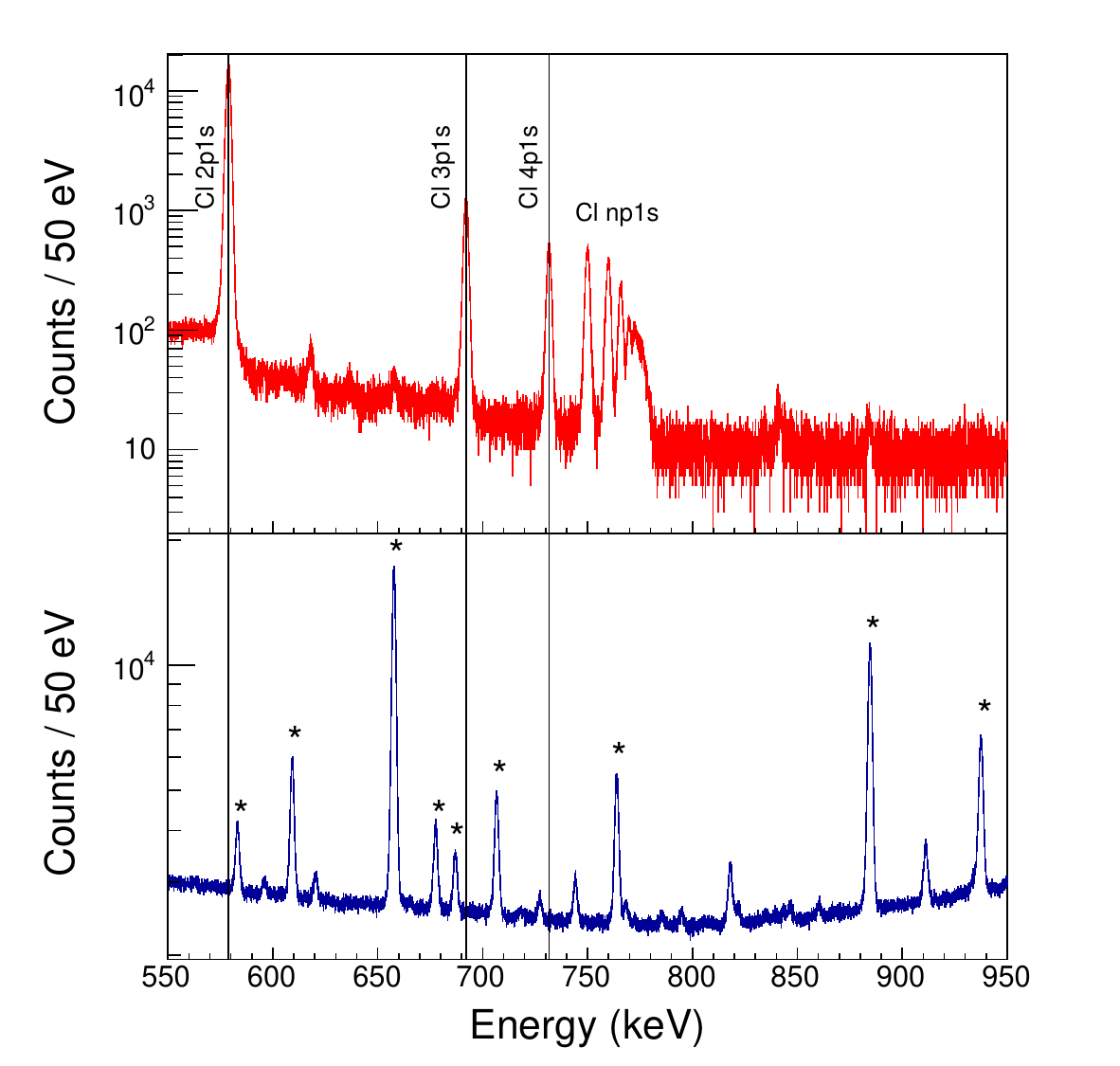}
        \caption{Prompt (top) and anticoincidence (bottom) spectra for the muonic x-ray measurements of \textsuperscript{35}Cl. The vertical lines mark the relevant peaks, the $np\to1s$ transition energies extracted in this work. The asterisk symbol labels the gamma-ray peaks used for the final energy calibration.}
        \label{fig:timecuts}
    \end{figure}

    Prompt spectra within $[-25; +25]\,\si{\nano\second}$ of an incoming muon revealed the muonic x-ray spectra, with the $np\to1s$ peaks lying between $550~\si{\kilo\eV}$ and $800~\si{\kilo\eV}$. Additionally, anticoincidence spectra that exclude all events within a window of $[-1, +3]\,\si{\micro\second}$ from any given muon provide a clean, continuous calibration spectrum. These are displayed in Fig.~\ref{fig:timecuts}, showing minimal contamination of calibration peaks in the prompt spectrum and vice versa.

    \begin{table*}
        \centering
        \caption{Obtained experimental values and corresponding uncertainty contributions for the transition energies and isotope shifts (see text for explanation of headers). Uncertainties should be interpreted as standard errors. Energies are given for the centroid of the fine-structure multiplets. The reduced $\chi^2$ is given for the fits ($\chi_{\nu, \text{fit}}^2$) and for the weighted average between the 19 detector crystals that were used ($\chi_{\nu, \text{av}}^2$).}
        \begin{ruledtabular}
        \begin{tabular}{cc|cccc|c|lcc}
            Isotope & Transition & $\sigma_{\text{stat+cal}}\,(\si{\eV})$ & $\sigma_{\text{bias}}\,(\si{\eV})$ & $\sigma_{\text{lit}}\,(\si{\eV})$ & $\sigma_{\text{f}}\,(\si{\eV})$ & $\sigma_{\text{exp}}\,(\si{\eV})$  & Energy  ($\si{\kilo \eV}$) & $\chi_{\nu, \text{fit}}^2$ & $\chi_{\nu, \text{av}}^2$ \\
            \hline
            \textsuperscript{35}Cl \rule{0pt}{2.5ex} 
            & $2p\to1s$ & 13.0 & 8.2 & 1.1 & 0.2 & 15.5 & 578.874(16) & 1.12 & 2.53 \\
            & $3p\to1s$ & \phantom{0}6.6  & 8.2 & 1.1 & 0.2 & 10.6 & 692.103(11) & 1.06 & 1.36 \\
            & $4p\to1s$ & \phantom{0}9.4  & 8.2 & 1.1 & 0.2 & 12.5 & 731.673(13) & 1.05 & 1.23 \\
            \hline
            \textsuperscript{37}Cl \rule{0pt}{2.5ex} 
            & $2p\to1s$ & \phantom{0}8.7  & 8.2 & 1.1 & 0.2 & 12.0 & 578.739(12) & 1.06 & 1.70 \\
            & $3p\to1s$ & 15.4 & 8.2 & 1.1 & 0.2 & 17.5 & 692.003(18) & 1.06 & 0.60 \\
            & $4p\to1s$ & 26.4 & 8.2 & 1.1 & 0.3 & 27.6 & 731.556(28) & 1.13 & 0.85 \\
            \hline
            \textsuperscript{37}Cl - \textsuperscript{35}Cl \rule{0pt}{2.5ex}  
            & $2p\to1s$ & 11.6 & / & / & 0.2 & 11.6 & $-$0.144(12) & / & 0.38 \\
            & $3p\to1s$ & 16.8 & / & / & 0.1 & 16.8 & $-$0.102(17) & / & 0.58 \\
            & $4p\to1s$ & 28.1 & / & / & 0.1 & 28.2 & $-$0.116(29) & / & 0.78
        \end{tabular}
        \end{ruledtabular}
        \label{tab:exp_uncertainties}
    \end{table*}

    The data were corrected for the gain drift of the germanium detectors using the anticoincidence spectra and finally calibrated to high precision in the range $[500; 1000]\,\si{\kilo\eV}$ for each detector crystal independently, using a hypermet line shape~\cite{campbell1997cautionary} with a coupled likelihood fit across all calibration peaks. To achieve the desired level of precision, an accurate energy calibration and uncertainty estimation are critical. First, the range on which the energy calibration was performed was set around the $np\to1s$ region. This reduced the number of major non-linearity jumps in the digitizer within the calibration range, thereby allowing it to be better described by a polynomial model. Secondly, the calibration was performed with orthogonal distance regression (ODR)~\cite{boggs1988computational} in combination with non-parametric bootstrapping~\cite{efron1994introduction, efron1981nonparametric}. The ODR accounts for the proper treatment of literature uncertainties, while the bootstrapping procedure provides a more accurate estimation of the fit uncertainty due to non-statistical effects (e.g, digitizer non-linearity). An example of the calibration residuals in one of the detectors is shown in the end matter.
     
\begin{figure}
        \centering
        \subfloat[]{\includegraphics[width = 1.0 \linewidth]{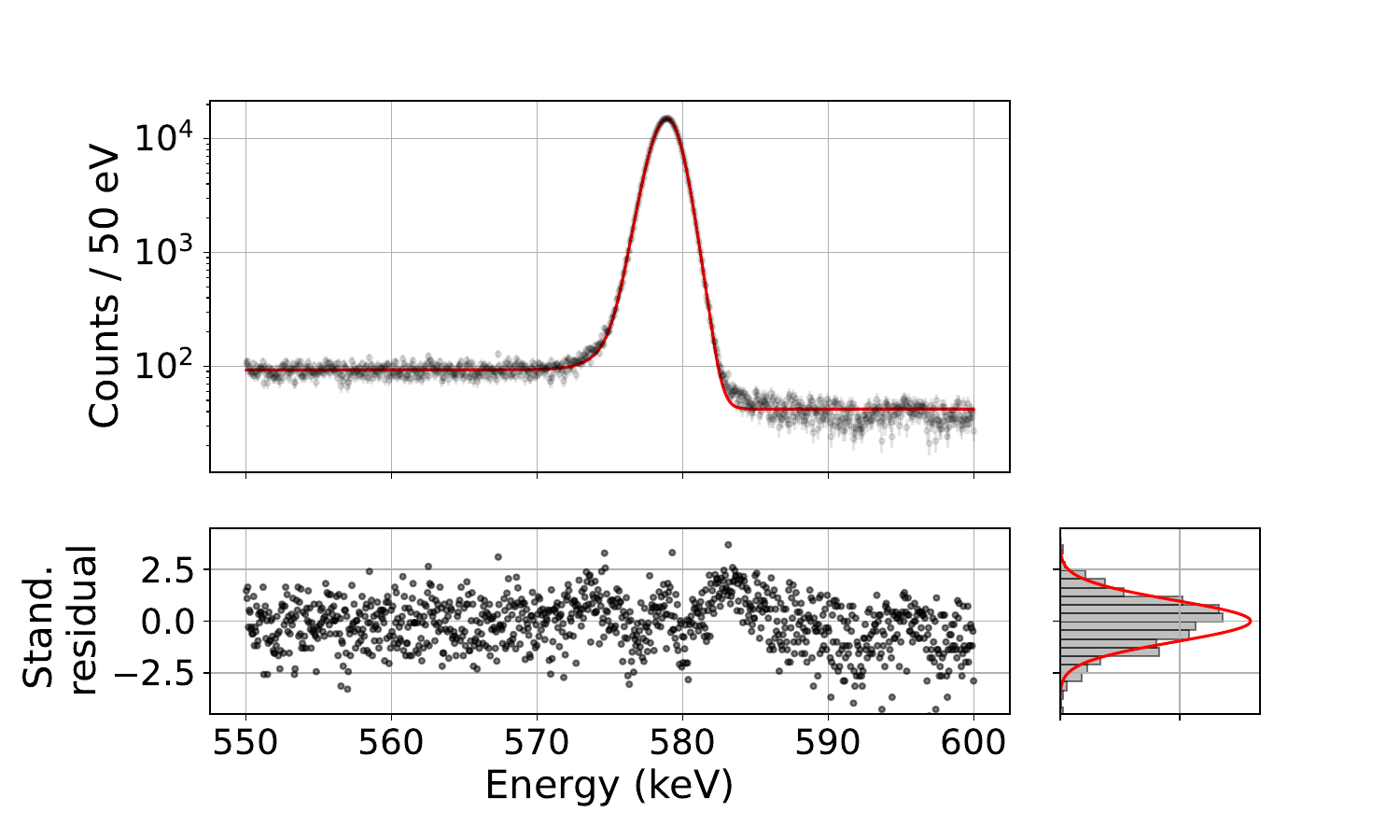}} \\
        \subfloat[]{\includegraphics[width = 1.0 \linewidth]{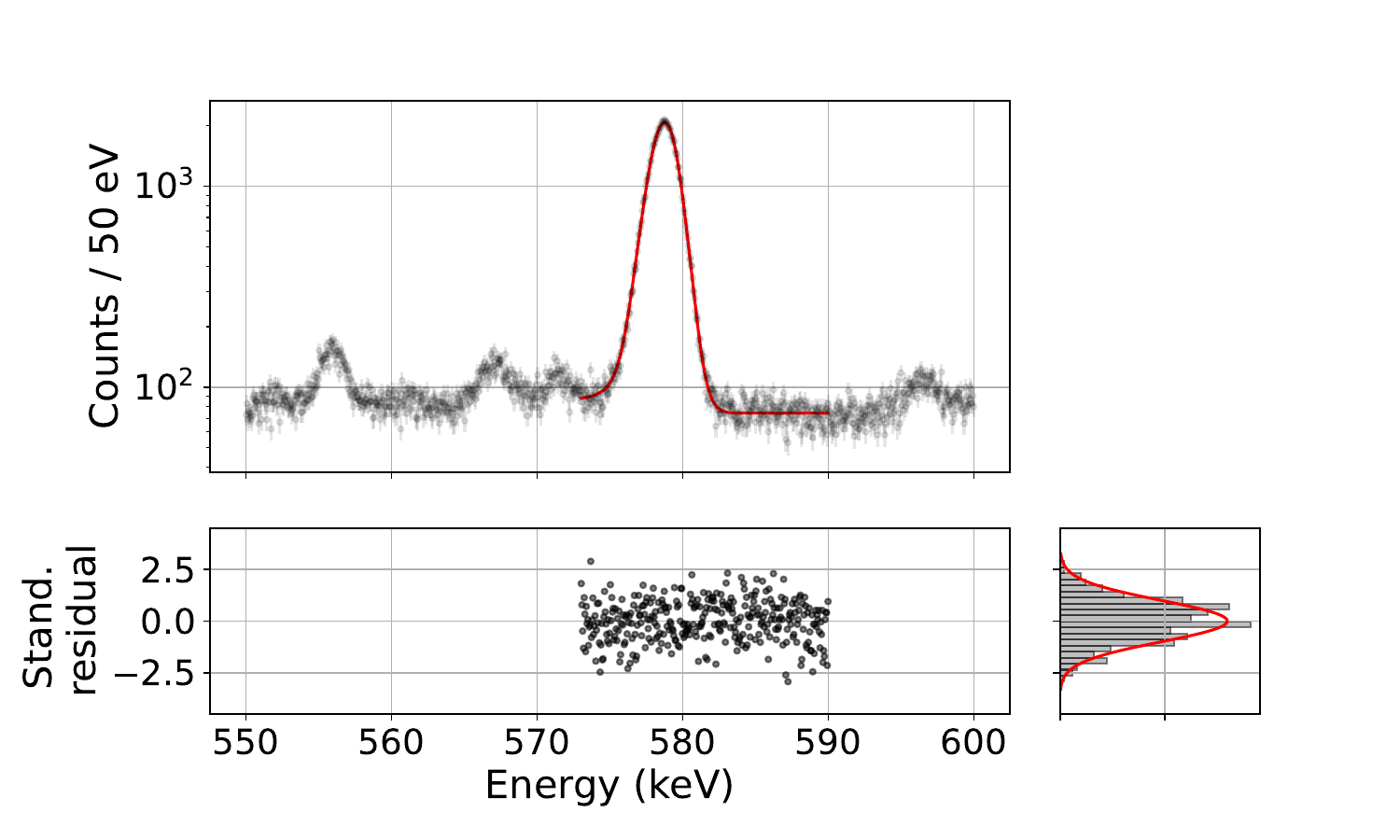}}
        \caption{Experimental spectrum of the $2p\to1s$ transition in (a) $^{35}$Cl and (b) $^{37}$Cl. The solid line corresponds to the sum of individual fits over all detectors. Standardized residuals are calculated using the total Poisson error on all detectors.}
        \label{fig:fit_2p1s}
    \end{figure}

    The energies of the $2p\to1s$, $3p\to1s$, and $4p\to1s$ x~rays were extracted with the same peak model as the calibration peaks, additionally accounting for the fine structure of the $2p$ state. This was done by means of a Bayesian inference fit using a peak doublet as a fitting model, with priors for the fine-structure splitting and intensity ratio rooted in QED calculations. In order to maximize accuracy and reliability, both the calibration and energy determination were performed in each detector separately. The statistical and calibration errors (which in themselves also have a statistical component) are added in quadrature before performing a weighted average over all of the detectors. The sum of data and fits for the $2p\to1s$ transition is shown in Fig.~\ref{fig:fit_2p1s}. For the case of $^{37}$Cl, the fit range was restricted to suppress the effect of minor nearby peaks originating from the Ag $nf\to3d$ transitions.

    Next, a minor correction was made to account for the isotopic impurity of the samples. Finally, the energies were shifted to account for the part of the transition energy carried away by the recoiling nucleus. We report the energy of the centroid of the $np\to1s$ transitions in Table~\ref{tab:exp_uncertainties}. Here, $\chi_{\nu, \text{fit}}^2$ and $\chi_{\nu, \text{av}}^2$ relate to the fitting process and the averaging over detectors. The former is calculated by summing the Bakers-Cousins $\chi^2$~\cite{baker1984clarification} of each detector and normalizing it to the total number of degrees of freedom.

    A large fraction of the uncertainty originates from the energy calibration. Since it is different for each detector, only the combined statistical and calibration uncertainty is reported ($\sigma_{\text{stat+cal}}$) after averaging over the different detectors. Additional sources of systematic uncertainties were identified and quantified. The first of these is an additional bias induced by the shared non-linearity of different digitizer channels. This effect was systematically tested across the calibration range and estimated to induce an additional uncertainty $\sigma_{\text{bias}}=8.2\,\si{\eV}$. The second contribution originates from the literature uncertainty of the most precisely known calibration line, contributing an additional $\sigma_{\text{lit}}=1.1\,\si{\eV}$. This effect is important, because several of the calibration lines were extracted in the same work, such that the uncertainty is partially correlated. The final contribution $\sigma_{\text{f}}$ to the uncertainty is caused by the uncertainty on the isotopic purity, which contributes minimally to the total uncertainty budget. A breakdown of these uncertainties is given in Table~\ref{tab:exp_uncertainties}, as well as the total experimental uncertainty $\sigma_{\text{exp}}$ calculated by means of quadrature addition. Other sources of systematic uncertainties, such as the omitted underlying hyperfine structure, were estimated to be $<0.1\,\si{\eV}$.

    %The uncertainty of these energies is dominated by the calibration uncertainty. Since this is different for each detector, only the combined statistical and calibration uncertainty is reported ($\sigma_{stat+cal}$) after averaging over the different detectors. However, additional systematic experimental uncertainties were identified. The shared non-linearity between digitizer channels, giving rise to a bias, was systematically tested across the calibration range and estimated to be $\sigma_{\text{bias}}=8.2\,\si{\eV}$. The uncertainty of the most precise calibration line contributed an additional $\sigma_{\text{lit}}=1.1\,\si{\eV}$. The uncertainty of the fractional isotopic purity leads to a minor systematic $\sigma_f$. Other sources of systematic effects, such as the omitted underlying hyperfine structure, were estimated to be $<0.1\,\si{\eV}$. The sources of uncertainties are given in Table~\ref{tab:exp_uncertainties} and were added in quadrature to produce the total experimental uncertainty $\sigma_{\text{exp}}$.

    From the data, isotope shifts between $^{35}$Cl and $^{37}$Cl could also be directly extracted. In this case, the bias and best calibration line uncertainties are strongly correlated between the different isotopes, such that they cancel out. This reduces the corresponding uncertainty, as presented in Table~\ref{tab:exp_uncertainties}. The averaging process was performed after calculating the energy differences within each detector separately, such that the isotope shifts are not necessarily equal to the differences between the quoted absolute energies (though they are well within one standard deviation). Additional details on experimental and analysis methods can be found in Ref.~\cite{heines2025muonic}.

\textit{Theory - }
    At the desired level of precision, existing theoretical QED frameworks developed for the lightest~\cite{pachucki2024comprehensive} or heavier~\cite{beyer2025relativistic,sun2025208} systems are not sufficient for the intermediate region considered. We therefore developed a hybrid approach to compute precise binding energies of muonic atoms and enable more accurate nuclear radius extractions for $3 \leq Z \lesssim 30$~\cite{2026-Hybrid}. Apart from QED effects, a major contribution to the uncertainty budget comes from the internal dynamic nuclear structure, commonly referred to as nuclear polarization (NP). A detailed description of the theoretical methods and the calculated level energies employed in this work is provided in a companion publication~\cite{2026-Hybrid}.

    The QED calculations require an assumed nuclear charge distribution, which introduces a significant level of model dependence. The muonic transition energies are only indirectly sensitive to the RMS charge radius, such that the \textit{shape} of the charge distribution acts as a systematic uncertainty. While in heavier systems it becomes possible to fit for several charge distribution parameters~\cite{sun2025208, beyer2025relativistic, saito2025muonic}, light and medium-mass systems lack the sensitivity to do this. To remedy the model dependence, the Barrett formalism was employed~\cite{barrett1970model, fricke2004nuclear}. Here, the Barrett moment $\langle r^{k} e^{-\alpha r}\rangle$ is introduced, where $k$ and $\alpha$ are determined from the calculated muon wavefunctions (see Table~\ref{tab:Barrett_parameters}). This moment is constructed to directly relate changes in the muonic transition energies to changes in the charge distribution, and as such is substantially less model dependent. Alongside this moment, the formalism introduces the Barrett equivalent radius $R_{k\alpha}$, which corresponds to the radius of a solid sphere with the same Barrett moment as the nucleus. The theoretical calculations were parametrized by fitting the transition energies (and isotope shifts) as a function of Barrett radius (and Barrett radius difference). The fit models and parameters of this parametrization are provided in the end matter, more details can be found in Ref.~\cite{heines2025muonic}.
    
    The Barrett recipe significantly reduces model dependence at the cost of a loss in direct physical relevance. This relevance is reintroduced later in the analysis using a correction factor ($V_2$), to calculate the RMS radius as $R_{\text{RMS}} = \frac{R_{k\alpha}}{V_2}$. This correction factor is extracted from a more accurate description of the shape of the monopole charge distribution of the respective nuclei~\cite{fricke2004nuclear, de1987nuclear}. Typical values lie a few per mill below $\sqrt{5/3}$, which corresponds to the translation from radius to RMS radius for a solid sphere. This correction is traditionally based on experimental measurements with elastic electron scattering. In many cases, including chlorine~\cite{briscoe1980elastic}, no such data are available of sufficient quality for an accurate determination. Here, we determined the $V_2$ correction factor from EDF calculations using the BSkG family of functionals~\cite{scamps2021skyrme, ryssens2022skyrme, grams2023skyrme, grams2025skyrme}. The uncertainty was estimated by comparing calculated values for $V_2$ with those obtained from high-precision electron-scattering data within the mass region relevant for this work. This indicated an average deviation of about 0.05\% (with a $\chi^2_{\nu}$ of 0.80), comparable to the spread between different precision experiments~\cite{ohayon2024critical}. This resulted in $V_{2, ^{35}\text{Cl}} = 1.28328(65)$ and $V_{2, ^{37}\text{Cl}} = 1.28382(65)$. In addition, $V_2$ correction factors of different isotopes of the same element showed a strong similarity and correlation. By comparing the difference in $V_2$ between the two isotopes in the EDF calculations and the experimental scattering data, the correlation between the $V_2$ of $^{35}$Cl and $^{37}$Cl is estimated to be $\sim$97\%.

\textit{Results - }
    By combining the experimental muonic x-ray energies from Table~\ref{tab:exp_uncertainties} with the theoretical calculations, the absolute charge radii of $^{35, 37}$Cl were extracted, as reported in Table~\ref{tab:result_rms}. Given the strong correlation of many experimental effects and theoretical calculations between the two isotopes, differences are extracted with increased precision. We report two sets of radii, one model-dependent but not relying on external input for the charge distribution and one relying on charge distributions calculated using the BSkG functionals. For the former, the charge distribution model was chosen to be a simple two-parameter Fermi distribution (2pF) with the skin thickness parameter set to $t=2.3~\si{\femto\meter} \pm 10\%$, a common assumption made in the past when no scattering data was available~\cite{fricke2004nuclear}. Additional details on how the theory and experiment were combined can be found in the end matter and Ref.~\cite{heines2025muonic}.

    \begin{table*}
        \centering
        \caption{Resulting RMS radii of \textsuperscript{35}Cl and \textsuperscript{37}Cl (in $\si{\femto\meter}$ and $\si{\femto\meter\squared}$). Additional zeroes were added to the literature values to emphasize the improvement in precision.}
        \begin{ruledtabular}
        \begin{tabular}{l lll}
            \multicolumn{1}{c}{} & \multicolumn{2}{c}{This work} & \multicolumn{1}{l}{Literature}~\cite{briscoe1980elastic} \\
            \cline{2-3}
            % & This work & This work & Literature \\
            Model & 2pF & BSkG4 & e\textsuperscript{-} scattering\\
            \hline
            $R^{35}$ \rule{0pt}{2.5ex} & 3.3323(48) & 3.3333(23) & \phantom{$-$}3.3880(170) \\ %3.3230(110) \\
            $R^{37}$ & 
            3.3446(48) & 3.3444(23) & \phantom{$-$}3.3840(170) \\ %3.3380(70) \\
            $R^{37} - R^{35}$ 
            & 0.01287(640) & 0.01163(95) & $-$0.00400(2400) \\%& 0.01500(1100) \\
            $\delta \langle r^2\rangle^{37, 35}$ 
            & 0.0860(430) & 0.0776(64) & $-$0.0300(1600) \\ %& 0.1030(700)
        \end{tabular}
        \end{ruledtabular}
        \label{tab:result_rms}
    \end{table*}

\begin{figure}[ht]
        \centering
        \includegraphics[width=0.99\linewidth]{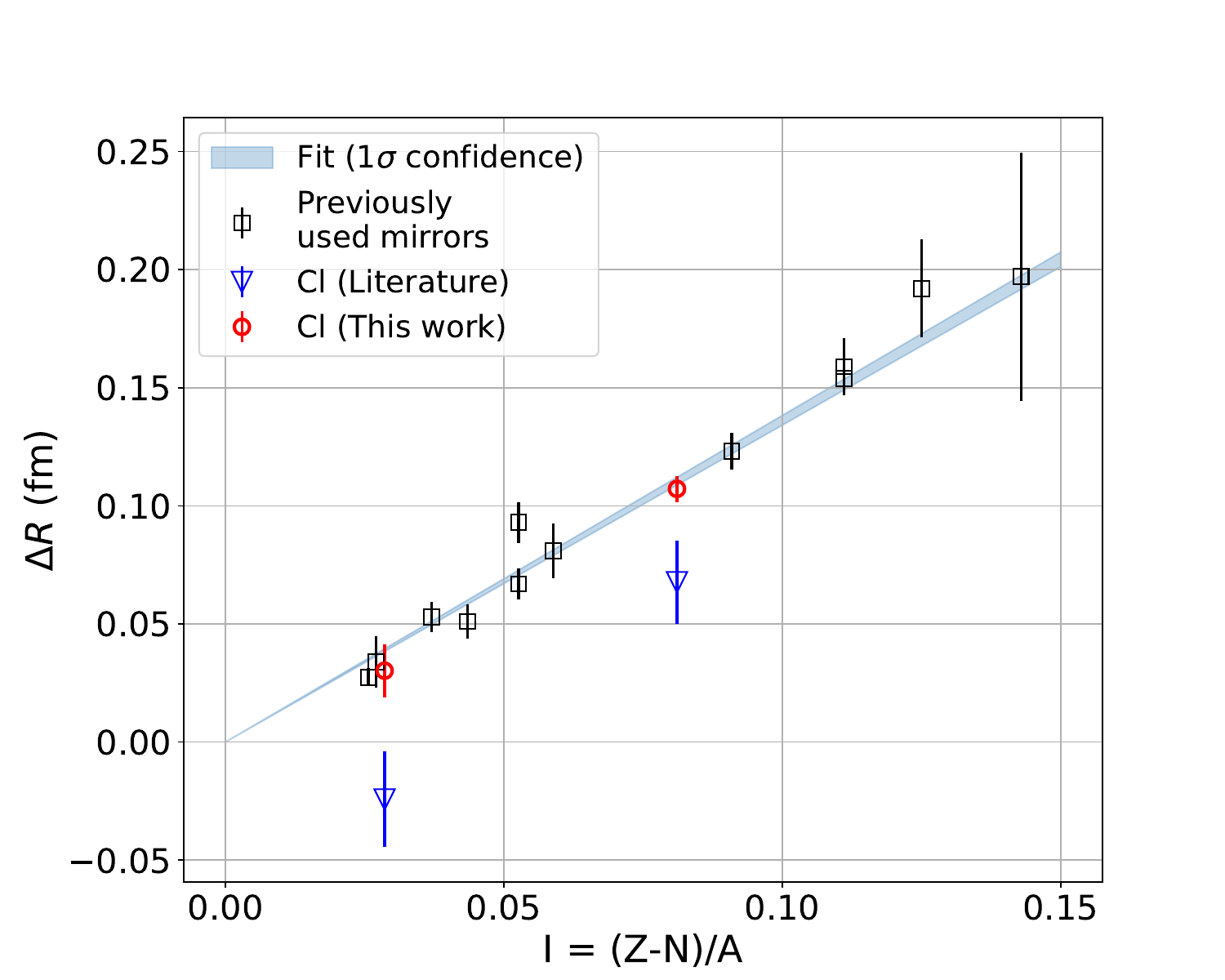}
        \caption{Mirror shift fit updated with chlorine radii from this work. Our measurements provide a substantial shift of the two mirror pairs involving stable chlorine isotopes, leading to better agreement with the global trend.}
        \label{fig:mirrorfit}
    \end{figure}

    Compared to the electron-scattering results from literature, the radii determined in this work are shifted by $3.2\,\sigma$ and $2.3\,\sigma$ for $^{35}$Cl and $^{37}$Cl, respectively, while the uncertainties are reduced by a factor of seven. Discrepancies between charge radii obtained from different experimental techniques have occasionally been reported in the past~\cite{de1987nuclear, angeli2013table} and are often attributed to differing sensitivities to the nuclear charge distribution or unaccounted systematic contributions (e.g., in Ref.~\cite{maisenbacher2026sub}). In the present comparison, the difference may be related to systematic effects in the literature electron scattering measurements~\cite{briscoe1980elastic}, for which only statistical uncertainties are reported. Additionally, our values are in much better agreement with the global behavior of mirror nuclei, shown in Fig.~\ref{fig:mirrorfit}. This model describes the behavior of the radius difference in a mirror pair $\Delta R$ as a function of the isospin asymmetry $I = (Z - N)/A$, which is expected to follow a proportional relationship~\cite{novario2023trends}. While the chlorine related points account for only two out of fourteen data points, the updated radii reduce the $\chi_\nu^2$ of the mirror shift fit by more than a factor of two (2.15 with literature radii compared to 1.01 with our updated values).

    %Furthermore, our values are in agreement with the global behavior of mirror nuclei shown in Fig.~\ref{fig:mirrorfit}. This model describes the behavior of the radius difference in a mirror pair $\Delta_I$ as a function of the isospin asymmetry $I = (Z - N)/A$. The behavior of this trend is expected to be approximately linear
    %due to neutron/proton skin effects 
    %In principle, one expects that the intercept would be zero, as a nucleus with no isospin asymmetry is its own mirror isotope and, as such, gives no difference in radius. It was also suggested that a fit of this trend could be used to estimate radii of isotopes without experimental data, if their corresponding mirror counterparts have been measured~\cite{ohayon2024critical}.

    Finally, the uncertainty on the difference in radius between the two isotopes is improved by a factor 25. With the improved precision, it can be observed that $^{37}$Cl is significantly larger than $^{35}$Cl, similarly to their isotonic equivalents~\cite{koszorus2021charge, miller2019proton, blaum2008nuclear}. This result is crucial for extracting charge radii of exotic chlorine isotopes from future isotope shift measurements by calibrating the ratio of isotope shift factors.

\textit{Conclusion - }
    There is renewed interest in absolute charge radii. Due to the various inputs and outstanding questions on the systematic uncertainties in muonic atoms, it is critical to reassess the radius extraction methods. With this analysis of the chlorine data we lay out an updated approach. By employing a large germanium detector array, a measurement could be performed on a few tens of milligrams of highly enriched chlorine. Additionally, improvements were made with more advanced data analysis procedures, the development of a theoretical framework suitable for medium-mass muonic atoms, and by employing nuclear theory to assess the nuclear shape correction. Finally, a more rigorous uncertainty evaluation was performed, and the correlation between observables was accounted for. 
    
    A high-precision germanium detector based measurement was performed on samples of $^{35}$Cl and $^{37}$Cl with high isotopic purity, for which the muonic $2p\to1s$, $3p\to1s$, and $4p\to1s$ energies were extracted. Our values show an improvement in the precision of the RMS radii by a factor of seven and a disagreement of $3.2\,\sigma$ and $2.3\,\sigma$ compared to the literature values~\cite{briscoe1980elastic} for the two isotopes. 
    
    These updated radii agree much better with the phenomenological mirror shift fit, adding confidence in our values. Furthermore, the muonic isotope shifts were used to extract a more accurate value for the difference in RMS radii and the differential mean-square radius, which is crucial for determining the radii of the entire chlorine chain in  upcoming measurements at online facilities. We demonstrated that advances in analysis methods, together with a more complete treatment of QED and nuclear-structure corrections, can improve the accuracy and precision of nuclear charge radii in the medium-mass region. We hope that the present investigation will trigger a revival of highly precise muonic x-ray experiments across the vast nuclear landscape.

\section*{Acknowledgments}
    The experiments were performed at the $\pi$E1 beamline of PSI. We would like to thank the accelerator and support groups for the excellent conditions. The germanium detector setup is shared with the MIXE project at PSI (\url{https://www.psi.ch/en/smus/giant-mixe}), which has greatly contributed to its construction, providing a fantastic platform for several muonic atom experiments taking place at PSI. This research used targets provided by the Center for Accelerator Target Science at Argonne National Laboratory, which is a DOE Office of Science User Facility and supported by the U.S. Department of Energy, Office of Nuclear Physics, under Award No.~DE-AC02-06CH11357; we further thank U.~K{\"o}ster for providing the target from ILL. 

    The authors acknowledge the following funding institutions: the Swiss National Science Foundation, Sinergia project “Deep$\mu$”, Grant: 193691 (MIXE); KU Leuven BOF under contract number C14/22/104; the European Research Council (ERC) through proposal No.~101088504 (NSHAPE); Fonds de la Recherche Scientifique (F.R.S.-FNRS), under project No.~F.4553.25 and the MIS project No.~40028446; the Romanian ministry of education and scientific research under project No.~PN~23~21~ 01~02, and contract No.~1/RO-CH/30.10.2025 within the Support Measure "International Research Cooperation Programme" 7F-11073; FWO Vlaanderen, through project G0G3121N (NSHAPE), and fellowship 11P6V24N (M.D.); the ETH Research Grant 22-2 ETH-023 (K.v.S.).; the Technion postdoctoral fellowship and Lady Davis Fellowship (S.R.). A.H. would like to thank the Slovak Research and Development Agency under contract No.~APVV-24-0516, and Slovak grant agency VEGA (contract No.~2/0175/24). M.G.~acknowledges support  by the Deutsche Forschungsgemeinschaft (DFG) - GO 2604/3-2, Projektnummer 495329596. N.S.O.~thanks the DFG (German Research Foundation) – Project-ID 273811115 – SFB 1225 ISOQUANT for funding. M.G., F.W. and R.P. are supported by the Cluster of Excellence “Precision Physics, Fundamental Interactions, and Structure of Matter” (PRISMA++ EXC 2118/2) funded by the German Research Foundation (DFG) within the German Excellence Strategy (Project ID 390831469).
    
    %Center of Excellence in High energy Physics and Astrophysics, Suranaree University of Technology (N.R.); 

    W.R.~is a Research Associate of the F.R.S.-FNRS (Belgium). Nuclear calculations were performed using computational resources from the Tier-1 supercomputer Lucia of the Fédération Wallonie-Bruxelles, infrastructure funded by the Walloon Region under the grant agreement No.~1117545, and the clusters Consortium des Équipements de Calcul Intensif (CÉCI), funded by F.R.S.-FNRS under Grant No.~2.5020.11 and by the Walloon Region.

\section*{Contribution statement}
    E.A.M. arranged the production of the \textsuperscript{110m}Ag calibration source; the experiment was performed by T.E.C., C.C., M.D., A.D., M.H., A.H., A.K., R.L., V.M., A.T., S.M.V. and K.v.S.; the analysis was performed by M.H., with input provided in discussions with T.E.C., C.C., M.D., A.D., O.E., A.K., R.L., B.O., W.W.M.M.P., R.P., S.M.V., K.v.S., F.W. and A.Z.; QED calculations and interpretation involved M.H., B.O., N.S.O., P.I. and S.R.; NP calculation and interpretation were performed by M.G., M.H., N.O. and I.V.; Barrett parameters were determined by K.A.B. and M.H.; the correction for the nuclear shape was evaluated by P.D., M.H., B.O. and W.R.; the manuscript was written by M.H. and T.E.C.; All co-authors reviewed the manuscript and were involved in practical discussions.

\section*{Competing interests}
    The authors declare no competing interests.

\section*{Data availability}
    Processed data sets are made available on Zenodo~\cite{dataset_Cl}.

\bibliography{biblio.bib}

\section*{End matter}
    \textit{Experimental - }
    The detector array used in this measurement consisted of a Miniball cluster detector~\cite{warr2013miniball}, a TIGRESS-type clover detector~\cite{scraggs2005tigress}, reverse electrode coaxial germanium detectors with relative efficiencies of 95\% (x1), and 70\% (x2), standard electrode coaxial germanium detectors with relative efficiencies of 50\% (x2), 58\% (x1), 75\% (x1), and 100\% (x2), and broad-energy germanium detectors (x3). A gain correction was performed assuming a Gaussian peak model using calibration lines from $^{60}$Co, \textsuperscript{110m}Ag and $^{133}$Ba. Beyond this, peaks were fitted with a hypermet line shape, given by 

    \begin{equation}\label{eq:hypermet}
        f(E) = N_{sig} \left[ f_G \times g(E) + f_T \times t(E) + s(E) \right],
    \end{equation}

    where

    \begin{eqnarray*}
        f_T =&& 1 - f_G, \\
        g(E) =&& \frac{1}{\sqrt{2\pi}\sigma} \exp{\left(-\frac{1}{2}\left[\frac{E - \mu}{\sigma}\right]^2\right)}, \\
        t(E) =&& \frac{1}{2 \beta} \exp{\left(\frac{E - \mu}{\beta} + \frac{\sigma^2}{2\beta^2}\right)} \\
        &&\times~\text{erfc}\left(\frac{E - \mu}{\sqrt{2} \sigma} + \frac{\sigma}{\sqrt{2}\beta}\right), \\
        s(E) =&& \frac{A}{2} \text{erfc}\left(\frac{E - \mu}{\sqrt{2}\sigma}\right).
    \end{eqnarray*}

    The three contributions are the ideal Gaussian detector response $g(E)$, a term accounting for incomplete charge collection $t(E)$ (primarily due to defects in the germanium crystal), and a step behavior $s(E)$ mainly induced by low-energy Compton scattering in the surrounding material~\cite{campbell1997cautionary, knoll2010radiation}. Using peak centroids extracted with this fitting model, a high-precision quadratic energy calibration was performed in a narrower energy range using the calibration peaks listed in Table~\ref{tab:calibration}, which includes natural background peaks from $^{208}$Tl and $^{214}$Bi.

    \begin{table}[ht]
        \centering
        \caption{Calibration lines used for the definitive energy calibration in this work. $^{208}$Tl and $^{214}$Bi are present in the natural Th decay chain. Values were taken from Ref.~\cite{be2004table}.}
        \begin{tabular}{c|c}
            Source & Energy ($\si{\kilo\eV}$) \\
            \hline
            $^{208}$Tl\rule{0pt}{3.0ex} & 583.187(2) \\
            $^{214}$Bi & 609.316(7) \\
            \textsuperscript{110m}Ag & 657.7600(11) \\
            \textsuperscript{110m}Ag & 677.6239(12) \\
            \textsuperscript{110m}Ag & 687.0114(18) \\
            \textsuperscript{110m}Ag & 706.6780(15) \\
            \textsuperscript{110m}Ag & 763.9452(17) \\
            \textsuperscript{110m}Ag & 884.6819(13) \\
            \textsuperscript{110m}Ag & 937.485(3) \\
        \end{tabular}
        \label{tab:calibration}
    \end{table}

\begin{figure}[h]
        \centering
        \subfloat[]{\includegraphics[width = 0.95 \linewidth]{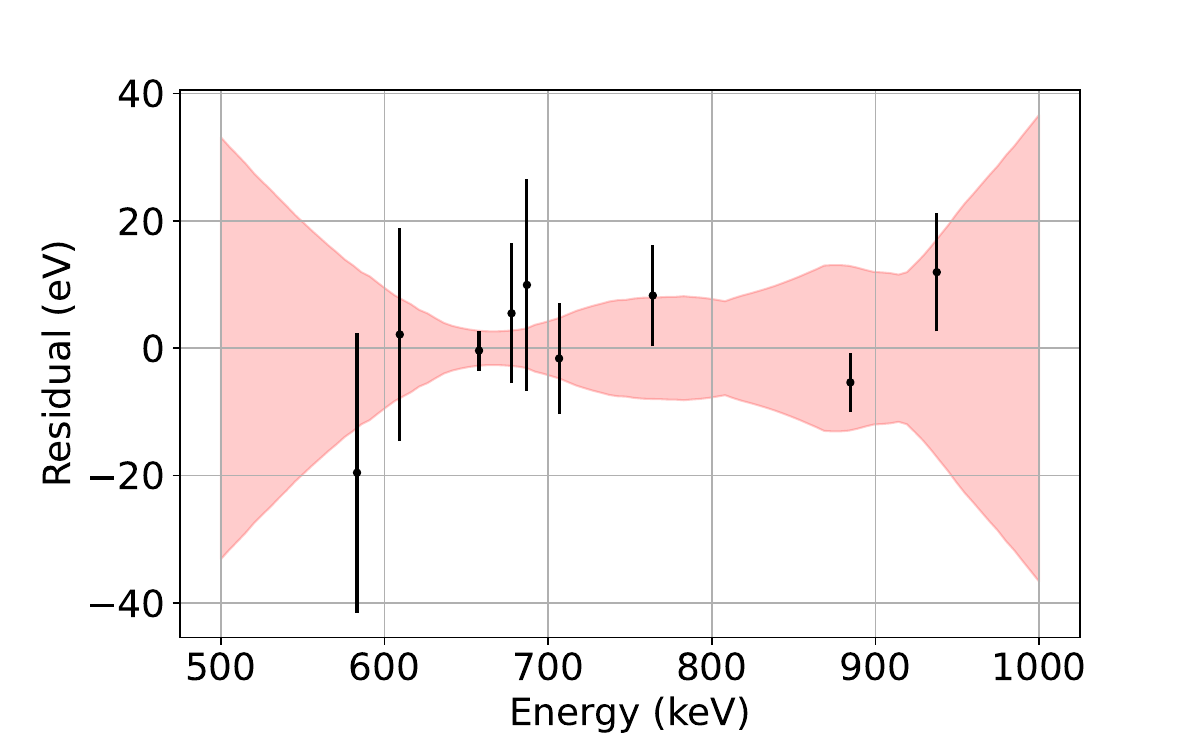}} \\
        \subfloat[]{\includegraphics[width = 0.95 \linewidth]{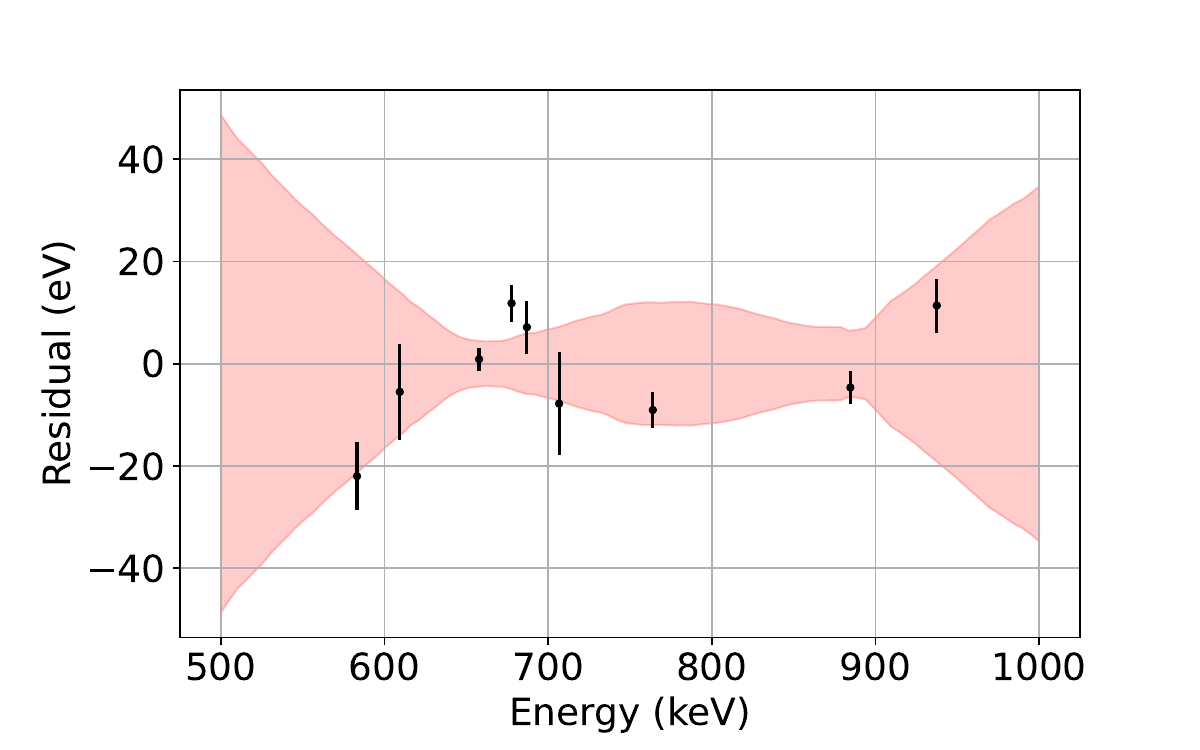}}
        \caption{Calibration residuals and estimated uncertainty for (a) statistics limited and (b) nonlinearity limited scenarios.}
        \label{fig:calib_resid}
    \end{figure}

    The calibration fit was performed using orthogonal distance regression to properly account for the statistical uncertainties as well as the uncertainty of literature values. Given that the calibration can become substantially limited by the non-linearity of the digitizer, the uncertainty was estimated using non-parametric bootstrapping within each individual detector. Given the relatively low number of data points, which can in extreme cases lead to an underestimation of the parameter variation, the uncertainty was estimated by taking the maximum of the statistical and bootstrap uncertainties at every energy. Fig.~\ref{fig:calib_resid} shows the calibration residuals and estimated uncertainty for a statistics-limited and nonlinearity-limited (long calibration measurement) case, showing realistic uncertainty estimates in both cases. More details on the analysis can be found in Ref.~\cite{heines2025muonic}.
    
    \textit{QED calculations - }
    A detailed description of the methodology behind the QED and NP calculations is provided in Ref.~\cite{2026-Hybrid}. Note that these calculations were made with a 2pF distribution with $t=2.3~\si{\femto\meter}$, such that they carry a substantial model dependence. While the comparison with experiments was only made after translating to Barrett equivalent radii, a parametrization was made of each QED contribution as a function of the RMS radius for the $2p\to1s$ transition. For this, a quadratic model was employed, recentred around $R_{\text{cen}} = 3.35~\si{\femto\meter}$

    \begin{equation}
        E_{np\to1s} = \alpha_0 + \alpha_1 (R - R_{\text{cen}}) + \alpha_2 (R - R_{\text{cen}})^2.
    \end{equation}
    
    Table~\ref{tab:QED_param_contrib} provides the parameters obtained from a least-squares fit for each contribution of the $2p\to1s$ transition. These contributions start from the energies obtained using the Coulomb and electronic Uehling potentials ($E(\mathrm{V_C+eVP_{11}})$). On top of this, perturbative corrections are included for the following effects: muonic Uehling potential ($\Delta E(\mathrm{\mu VP_{11}})$), hadronic VP ($\Delta E(\mathrm{hVP_{11}})$), Källén-Sabry VP ($\Delta E(\mathrm{eVP^2_{11}})$), Wichmann-Kroll VP ($\Delta E(\mathrm{eVP_{13}})$), self energy ($\Delta E(\mathrm{SE})$), interlinked self energy with electronic VP ($\Delta E(\mathrm{SE-eVP})$), leading and second order recoil ($\Delta E_{\mathrm{rec}}^{(1)}$, $\Delta E_{\mathrm{rec}}^{(2)}$), recoil corrections to the self energy and vacuum polarization ($\Delta E(\mathrm{SE_{rec}})$, $\Delta E(\mathrm{VP_{rec}})$), electron screening ($\Delta E_{\mathrm{screening}}$), and nuclear polarization ($\Delta E(\mathrm{NP})$). Table~\ref{tab:QED_param_np1s} provides a similar parametrization for the sum of all contributions for all studied $np\to1s$ transitions.

    \begin{table}[ht]
        \centering
        \begin{ruledtabular}
        \begin{tabular}{l|l|rrr}
            & Contribution & $\alpha_0$\phantom{(000)} & $\alpha_1$ & $\alpha_2$ \\
            % & Contribution & ($\si{\eV}$)\phantom{(000)} & ($\si{\eV\per\femto\meter}$) & ($\si{\eV\per\femto\meter\squared}$) \\
            \hline
            Both & $E(\mathrm{V_C+eVP_{11}})$ 
            & 580~237.2\phantom{(000)} & $-$17~731 & $-$920 \\
            & $\Delta E(\mathrm{\mu VP_{11}})$
            & 8.0\phantom{(000)} & $-$ 2 \\
            & $\Delta E(\mathrm{hVP_{11}})$
            & 5.7\phantom{(000)} & $-$1 \\
            & $\Delta E(\mathrm{eVP^2_{11}})$
            & $-$2.0\phantom{(000)} \\
            & $\Delta E(\mathrm{eVP_{13}})$
            & 34.0\phantom{(000)} & $-$3 \\
            & $\Delta E(\mathrm{SE})$
            & $-$150.3(40)\phantom{0} & 29 \\
            & $\Delta E(\mathrm{SE-eVP})$
            & $-$2.0(20)\phantom{0} \\
            & $\Delta E(\mathrm{SE_{rec}})$
            & 1.3\phantom{(000)} \\
            & $\Delta E_{\mathrm{rec}}^{(2)}$ 
            & 6.9\phantom{(000)} \\
            & $\Delta E_{\mathrm{screening}}$ 
            & $-$0.4\phantom{(000)} \\
            \hline
            $^{35}$Cl\rule{0pt}{2.5ex}
            & $\Delta E_{\mathrm{rec}}^{(1)}$
            & $-$1~674.4\phantom{(000)} & 144 \\
            & $\Delta E(\mathrm{VP_{rec}})$ & 16.8\phantom{(000)} \\
            & $\Delta E(\mathrm{NP})$ 
            & 115.2(250) \\
            $^{37}$Cl\rule{0pt}{2.5ex}
            & $\Delta E_{\mathrm{rec}}^{(1)}$
            & $-$1~543.9\phantom{(000)} & 136 \\
            & $\Delta E(\mathrm{VP_{rec}})$ & 15.9\phantom{(000)} \\
            & $\Delta E(\mathrm{NP})$ 
            & 112.0(250)
        \end{tabular}
        \end{ruledtabular}
        \caption{Parametrization of the calculated QED contributions for the $2p\to1s$ transition in chlorine. Quoted values are in $\si{\eV}$, $\si{\eV\per\femto\meter}$ and $\si{\eV\per\femto\meter\squared}$ for $\alpha_0$, $\alpha_1$ and $\alpha_2$, respectively. The values were rounded to ensure a rounding error of $0.1~\si{\eV}$ in the calculated RMS range. For details on the uncertainties, see Ref.~\cite{2026-Hybrid}.}
        \label{tab:QED_param_contrib}
    \end{table}

    \begin{table}[ht]
        \centering
        \begin{ruledtabular}
        \begin{tabular}{l|c|ccc}
            Isotope & Transition & $\alpha_0$\phantom{(000)} & $\alpha_1$ & $\alpha_2$ \\
            \hline
            $^{35}$Cl\rule{0pt}{2.5ex} 
            & $2p\to1s$ & 578~562.4 & $-$17~563 & $-$920 \\
            & $3p\to1s$ & 691~788.2 & $-$17~572 & $-$920 \\
            & $4p\to1s$ & 731~375.3 & $-$17~574 & $-$920 \\
            \hline
            $^{37}$Cl\rule{0pt}{2.5ex}
            & $2p\to1s$ & 578~650.6 & $-$17~571 & $-$920 \\
            & $3p\to1s$ & 691~896.3 & $-$17~579 & $-$920 \\
            & $4p\to1s$ & 731~490.3 & $-$17~582 & $-$920
        \end{tabular}
        \end{ruledtabular}
        \caption{Parametrization of the calculated $np\to1s$ transition energy in chlorine. Quoted values are in $\si{\eV}$, $\si{\eV\per\femto\meter}$ and $\si{\eV\per\femto\meter\squared}$ for $\alpha_0$, $\alpha_1$ and $\alpha_2$, respectively. Values were rounded to ensure a rounding error of $0.1~\si{\eV}$ in the calculated RMS range.}
        \label{tab:QED_param_np1s}
    \end{table}

    \textit{Barrett radii - }
    The parameters ($k$ and $\alpha$) for which the Barrett moment becomes insensitive to the shape of the monopole charge distribution can be obtained by fitting the electrostatic potential induced by the muon~\cite{fricke2004nuclear}. This potential was obtained by plugging the muon wavefunction, extracted with a simplified QED calculation using a realistic nuclear charge distribution, into the Poisson equation. This potential was then shifted such that $V(0)=0$ and fitted with the model $f(r) = B r^k e^{-\alpha r}$. For this fit, the data were weighted by $r^2 \rho(r)$, in order to weigh with the overlap with the nuclear wavefunction. The extracted parameters (given in Table~\ref{tab:Barrett_parameters}) are mostly similar across the different transitions (as they are dominated by the $1s$ orbital), such that their average value could be used. Tests showed a sub-$\si{\eV}$ residual model dependence, such that it could be safely ignored.

    \begin{table}[ht]
        \centering
        \caption{Extracted Barrett parameters for the $np\to1s$ transitions in Cl. Uncertainties on the average represent the largest difference with an individual transition.}
        \begin{ruledtabular}            
        \begin{tabular}{cc|ccc|c}
            Isotope & Parameter & $2p1s$ & $3p1s$ & $4p1s$ & Average \\
                \hline
                $^{35}$Cl\rule{0pt}{2.5ex} 
                    & $k$      & 2.0941 & 2.0936 & 2.0934 & 2.0937(4) \\
                    & $\alpha$ & 0.0561 & 0.0559 & 0.0558 & 0.0559(2) \\
                \hline
                $^{37}$Cl\rule{0pt}{2.5ex} 
                    & $k$        & 2.0944 & 2.0939 & 2.0937 & 2.0940(4) \\
                    & $\alpha$   & 0.0560 & 0.0557 & 0.0557 & 0.0558(2) \\
        \end{tabular}
        \end{ruledtabular}
        \label{tab:Barrett_parameters}
    \end{table}

    To extract the Barrett radii, the QED calculations were parametrized using

    \begin{equation}
        R_{k \alpha} = a_0 + a_1 (R_{k\alpha} - R_{k\alpha}^{\text{cen}}) + a_2 (R_{k \alpha} - R_{k\alpha}^{\text{cen}})^2,
    \end{equation}

    where $R_{k\alpha}^{\text{cen}} = 4.3~\si{\femto\meter}$ was set to reduce correlation between the fit parameters. This model described the calculations within $<0.1~\si{\eV}$. The fit parameters are given in Table~\ref{tab:radii_QED_Rka}. \\
    
    \begin{table}[ht]
        \centering
        \caption{Fit parameters for the energy of the $2p1s$, $3p1s$, and $4p1s$ lines. Uncertainties on the fit parameters are always smaller or similar to the rounding error.}
        \begin{ruledtabular}
        \begin{tabular}{cc|ccc}
            Isotope & Transition 
            & $a_0~(\si{\kilo\eV})$ 
            & $a_1~(\si{\kilo\eV\per\femto\meter})$ 
            & $a_2~(\si{\eV\per\femto\meter\squared})$ \\
            \hline
            $^{35}$Cl\rule{0pt}{2.5ex} 
                & $2p\to1s$ & 578.56456 & $-$13.623 & $-$551 \\
                & $3p\to1s$ & 691.79032 & $-$13.629 & $-$553 \\
                & $4p\to1s$ & 731.37743 & $-$13.631 & $-$553 \\[1mm]
            \hline
            $^{37}$Cl\rule{0pt}{2.5ex} 
                & $2p\to1s$ & 578.65413 & $-$13.629 & $-$551 \\
                & $3p\to1s$ & 691.89980 & $-$13.635 & $-$553 \\
                & $4p\to1s$ & 731.49381 & $-$13.637 & $-$553
        \end{tabular}
        \end{ruledtabular}
        \label{tab:radii_QED_Rka}
    \end{table}
    
    Similarly, the Barrett radius difference was parametrized as

    \begin{align}
        \Delta R_{k \alpha} &= b_0 + b_1 \Delta R_{k\alpha} + b_2 (R_{k \alpha, A'} - R_{\text{cen}}) \nonumber \\
        &+ b_3 (\Delta R_{k\alpha})^2 + b_4 (R_{k \alpha, A'} - R_{\text{cen}}) \Delta R_{k \alpha}.
    \end{align}

    Again, this model described the calculation within $<0.1~\si{\eV}$. The fit parameters are given in Table~\ref{tab:radii_QED_dRka}. Omitting the cross term described by $b_4$ leads to approximately $5~\si{\eV}$ residual variance.\\

    \begin{table}[ht]
        \centering
        \caption{Fit parameters for the description of the muonic isotope shifts as a function of Barrett radius differences. Uncertainties on the fit parameters are always smaller or similar to the rounding error.}
        \begin{ruledtabular}
        \begin{tabular}{c|lll}
            Transition & $2p\to1s$ & $3p\to1s$ & $4p\to1s$ \\
            \hline
            $b_0~(\si{\eV})$ &
            \phantom{$-$}89.57 & \phantom{$-$}109.47 & \phantom{$-$}116.38 \\
            $b_1~(\si{\kilo\eV\per\femto\meter})$ &
            $-$13.6227 & $-$13.6293 & $-$13.6313 \\
            $b_2~(\si{\eV\per\femto\meter})$ &
            $-$6.09 & $-$6.11 & $-$6.09 \\
            $b_3~(\si{\eV\per\femto\meter\squared})$ &
            \phantom{$-$}551 & \phantom{$-$}553 & \phantom{$-$}553 \\
            $b_4~(\si{\kilo\eV\per\femto\meter\squared})$ &
            $-$1.102 & $-$1.105 & $-$1.106 \\
        \end{tabular}
        \end{ruledtabular}
        \label{tab:radii_QED_dRka}
    \end{table}

    The extracted Barrett radii and Barrett radius difference are given in Table~\ref{tab:radii_Barrett_values}. Given that the NP and QED uncertainties are dominated by the $1s$ level (acting on all measured transitions), the averaging across transitions was performed using the inverse of the experimental variance as weights. The $\chi_\nu^2$ of the averaging process are very close to the medians of the statistical $\chi_\nu^2$ distributions, which provides an indication that the experimental error estimation is accurate. \\

    \textit{Results - }
    To extract RMS charge radii, the Barrett radii were combined with $V_2$ from EDF calculations. A breakdown of the considered uncertainties on the resulting values is given in Table~\ref{tab:uncertainty_breakdown}. At the present precision, the absolute radii are limited by the knowledge of the NP and $V_2$ corrections, while the differences are about equally restricted by experimental, NP, and $V_2$ uncertainties.\\

    \begin{table}[ht]
        \centering
        \caption{Uncertainty breakdown for the extracted radii, quoted in $10^{-3}\si{\femto\meter}$ and $10^{-3}~\si{\femto\meter\squared}$. Values should be interpreted as standard errors. Energies are given for the centroid of the fine-structure multiplets.}
        \begin{ruledtabular}            
        \begin{tabular}{cc|cccc|c}
            Isotope & model & $\sigma_{\text{Exp}}$ & $\sigma_{\text{NP}}$ & $\sigma_{\text{QED}}$ & $\sigma_{V_2}$ & $\sigma_{\text{tot}}$ \\
            \hline
            $R^{35}$\rule{0pt}{2.5ex}
            & 2pF   & 0.41 & 1.43 & 0.23 & 4.49 & 4.74 \\
            & BSkG4 & 0.41 & 1.43 & 0.23 & 1.69 & 2.27 \\
            \hline
            $R^{37}$\rule{0pt}{2.5ex}
            & 2pF   & 0.53 & 1.38 & 0.23 & 4.46 & 4.70 \\
            & BSkG4 & 0.53 & 1.38 & 0.23 & 1.69 & 2.26 \\
            \hline
            $R^{37} - R^{35}$\rule{0pt}{2.5ex}
            & 2pF   & 0.52 & 0.68 & / & 6.33 & 6.39 \\
            & BSkG4 & 0.52 & 0.68 & / & 0.41 & 0.95 \\
            \hline
            $\delta \langle r^2 \rangle^{37, 35}$\rule{0pt}{2.5ex}
            & 2pF   & 3.5 & 4.5 & / & 42.2 & 42.6 \\
            & BSkG4 & 3.5 & 4.5 & / & 2.8 & 6.4
        \end{tabular}
        \end{ruledtabular}
        \label{tab:uncertainty_breakdown}
    \end{table}

    Finally, the fit parameters for the mirror shift fit (Fig.~\ref{fig:mirrorfit}) under different fitting conditions are given in Table~\ref{tab:mirrorfit}.

    \begin{table}[ht]
        \centering
        \caption{Results from the proportional mirror shift fit ($\Delta R = c_1 I$) performed under different conditions, see Fig.~\ref{fig:mirrorfit}.}
        \begin{ruledtabular}
        \begin{tabular}{c|c|c|c}
            Parameter        & Excluding Cl & Literature Cl & This work \\
            \hline
            $\nu$            & 11          & 13          & 13          \\
            $\chi_\nu^2$     & 1.08        & 2.15        & 1.01        \\
            $c_1$            & 1.380(36)   & 1.362(50)   & 1.367(32)   \\
          % \hline
          % $\nu$            & 10          & 12          & 12          \\
          % $\chi_\nu^2$     & 0.98        & 2.06        & 0.90        \\
          % $d_0 (10^{-3})$  & $-$6.4(4.4) & $-$8.0(6.3) & $-$6.7(4.2) \\
          % $d_1$            & 1.465(68)   & 1.468(98)   & 1.457(63)   \\
          % Corr($d_0, d_1$) & -86.8\%     & -86.9\%     & -87.9\%
      \end{tabular}
      \end{ruledtabular}
      \label{tab:mirrorfit}
    \end{table}

    \begin{table*}[h]
        \centering
        \caption[Extracted Barrett radii]{Extracted Barrett radii (in $\si{\femto\meter}$) from this work. The round brackets (~), square brackets [~], and curly brackets \{~\} represent the uncertainty originating from the experimental energy determination, the NP correction, and QED. $\chi_{\nu, \text{av}}^2$ denotes the reduced $\chi^2$ of the average over transitions.}
        \begin{ruledtabular}
        \begin{tabular}{c|lll|lc}
            Isotope & $2p\to1s$ & $3p\to1s$ & $4p\to1s$ & Average & $\chi_{\nu, \text{av}}^2$ \\
            \hline
            $^{35}$Cl\rule{0pt}{3.0ex} 
            & 4.2772(12)[18]\{3\} & 4.2770(8)[19]\{3\} & 4.2783(10)[19]\{3\} 
            & 4.2775(6)[19]\{3\} & 0.58 \\
            $^{37}$Cl
            & 4.2938(9)[18]\{3\} & 4.2924(13)[18]\{3\} & 4.2954(21)[18]\{3\} 
            & 4.2936(7)[18]\{3\} & 0.87 \\[1mm]
            
            $^{37}$Cl $-$ $^{35}$Cl\rule{0pt}{3.0ex}
            & 0.0172(9)[9] & 0.0156(13)[9] & 0.0171(21)[9] 
            & 0.0167(7)[9] & 0.61
        \end{tabular}
        \end{ruledtabular}
        \label{tab:radii_Barrett_values}
    \end{table*}

\end{document}